\documentstyle[pra,aps,epsf,floats]{revtex}

\def\be{\begin{equation}}
\def\ee{\end{equation}}
\def\bea{\begin{eqnarray}}
\def\eea{\end{eqnarray}}
\begin{document}

\draft

\twocolumn[\hsize\textwidth\columnwidth\hsize\csname@twocolumnfalse\endcsname

\title{On the observation of decoherence with a movable mirror}
\author{R.Folman$^{1,2}$, J.Schmiedmayer$^2$, H.Ritsch$^3$ and D.Vitali$^4$}
\address{$^1$Institut f\"{u}r Experimental Physik, Universit\"{a}t Innsbruck,
Technikerstra{\ss} e 25, A--6020 Innsbruck, Austria\\
$^2$Physikalisches Institut,
    Universit\"at Heidelberg, D-69120 Heidelberg Germany\\
$^3$Institut f\"{u}r Theoretische Physik, Universit\"{a}t
Innsbruck, Technikerstra{\ss} e 25, A--6020 Innsbruck, Austria\\
$^4$Dip. di Matematica e Fisica, and Unit\`a INFM, Universit\`a di
Camerino, I-62032 Camerino, Italy}

\date{\today}

\maketitle

\begin{abstract}
Recently it has been proposed to use parity as a measure of the
mechanism behind decoherence or the transformation from quantum to
classical. Here, we show that the proposed experiment is more
feasible than previously thought, as even an initial thermal
state would exhibit the hypothesized symmetry breaking.
\end{abstract}


\vskip1pc

]


"The ball is in box1 {\em and} the ball is in box2" versus "the
ball is in box1 {\em or} the ball is in box2", is well known to
describe the difference between the quantum world and our
classical reality. The transition from a position superposition to
a well localized state is clearly a spatial symmetry breaking
transition (let the symmetry axis lie half way between the boxes
and perpendicular to the line that joins them). One should note
that although a {\em symmetric} density matrix
in position space $\rho_f$ is formed in the place of that
describing the initial superposition $\rho_i$, symmetry has been
broken as may be revealed by the different outcomes of applying
the parity operator $P$ onto the above two density matrices, e.g.
in a spatial two dimensional space of $|R>$ and $|L>$

\[%
\begin{array}
[c]{cc}%
\rho_{i}=\frac{1}{2}\left(
\begin{array}
[c]{ll}%
1 & \pm 1\\
\pm 1 & 1
\end{array}
\right)  & \rho_{f}=\frac{1}{2}\left(
\begin{array}
[c]{lll}%
1 & 0\\
0 & 1
\end{array}
\right)  ,
\end{array}
\]

where

\[%
\begin{array}
[c]{cc}%
P=\left(
\begin{array}
[c]{ll}%
0 & 1\\
1 & 0
\end{array}
\right)  ,
\end{array}
\]

and where $<P_f>=\frac{Tr(P\rho_f)}{Tr(\rho_f)}=0$ while
$<P_i>=\pm 1$.


The above symmetry breaking has in the past received many different
names, examples being `an emerging element of reality' or 'localization'  
and more recently 'dephasing' and 'decoherence'. In attempting 
to quantify this process of transition into what is sometimes referred to as
'classical reality' or a 'classical state', 
perhaps the notion of a statistical mixture is the more 
convenient description, where the declining magnitudes of the off diagonal
elements of the density matrix are the measure of the decoherence. 
There has been
much debate on what is the origin of this process and whether or not it is
independent of us and our measurements \cite{dec}. 
In order to resolve this issue one has to measure the process.
However, until now the question remains, 
how can one observe this process in
x space without initiating it? Let us elaborate: 
Observing a quantum system needs coupling to an external detection system.
In general, this influences the dynamics and causes symmetry breaking.
Hence, it is
clear that by observing a quantum system, we couple
our detectors to it and consequently allow its symmetry to
break and thus mask the very process we wish to observe, namely,
the emmergence of a statistical mixture independent of
our measurement. This may be formally described as

\bea
P^2+W^2=1
\label{know1}
\eea

where $P$ is a particle measure, $W$ is a wave measure which is simply
the visibility of an interference pattern, 
and they both take values between $0$ and $1$
\cite{greenberger}.

The particle measure is usually just the ability to predict
which path a particle will take i.e. to predict the outcome of a
"box1 {\em or} box2" experiment. However,
it may also be referred to as the knowledge $K$ possibly obtained in a 
which path information measurement \cite{englert}, and the above equality
is then re-written as

\bea
K^2+V^2=1
\label{know3}
\eea

where $V$ is now the visibility.

The answer proposed by Folman et al. \cite{folman} to this
experimental dilemma is that one searches for symmetry breaking
through parity eigen state measurements which do not reveal which
path information i.e. the coupling to the measurement apparatus is
facilitated through a parity conserving pointer basis set. 
Namely, that another measurement basis exists
with which the emmergence of a 'classical state' in the form of a statistical
mixture can be observed, which is different from the usual 'which path' basis
which is simply the previously introduced $|R>$ and $|L>$.
This basis, the parity basis, will not, as will be seen in the following,
couple to the system in a way that would mask other processes causing spatial
decoherence.

It is perhaps interesting to note that relative to the usual basis, the parity
basis in not intuitive as a measure of a 'classical state' in the sense that 
it does not observe 'particles' - the intuitive core of a 'classical state'. 
Hence, people have usually considered that measuring the appearence of a 
'classical state' and looking for 'particles' are one of the same.
As the notion of 'particles' is only valid in the `which path' basis, 
the latter is the basis commonly used, as is also evident from equations 
\ref{know1} and \ref{know3}. 

Consequently, re-defining the `particle measure' $P$ as a 
'classical state' measure
i.e. the measure of a statistical mixture and not more, 
we find that we can be sensitive
to $P$ while not suppressing the value of $W$ which is the measure of the
quantum state.  
%
%
%
We should note,
that according to the usual definition of $P$ ,
the above equalities remain of course unbroken also in our case because the
'which path' predictability/knowledge remains zero. In the sense that a 
'particle' is just a euphemism for 'classical state' or statistical mixture, 
we feel that the mathematical definition originally given to $P$ does not 
encompass fully the stated function of a 'particle measure'.

To give a simple example of the procedure proposed in \cite{folman}, 
it is well known that once a chiral
molecule has localized and hence created an element of reality in
the form of its well defined handedness, it is no longer in the parity
eigen state, displaying once again the symmetry border
between classical and quantum. As measuring parity does not
localize the molecule, observing a parity change of an isolated
molecule could only mean that "something else" localized it.
Indeed, Folman et al. have shown how parity eigen state
measurements could be used to investigate different models of
decoherence. On the other hand, if one tries to observe the
molecule's evolution from a parity eigen state into a well defined
handedness by measuring its handedness, he will initiate that
evolution himself.

In this letter we would like to extend the scope of the
previously mentioned experimental proposal by showing that even
if the initial quantum system is in a thermal state, the expected
symmetry breaking signal would still be observable. This has an
important consequence as extreme cooling of macroscopic objects
to their ground state, or difficult verification of an initial
pure state, would not be needed. Consequently, the proposed
experiment is much more feasible than previously thought.

Before beginning, let us briefly remind ourselves of the
experimental set up proposed \cite{folman}. We present it in
figure 1: A closed loop triangle interferometer consisting of one beam 
splitter rather than the usual Mach-Zehnder double beam splitter set up,
is turned into an open interferometer (two separate non overlaping optical 
paths) by introducing a double sided mirror into the set up. The mirror plane
coincides with the plane of the beam splitter, which constitutes the spatial
symmetry axis of the set up. The beam splitter, including a phase shifter
which compensates for the phase difference between the transmitted and 
reflected beams, prepares an initial symmetric photon. Together with the 
two detectors, it also acts as a measurement with parity eigen states
on the outgoing photon. 

After the incoming photon interacts with the mirror, it may
excite it or not, where the chance of the latter occurrence is
just the Debye-Waller factor $P_{0\rightarrow 0}$. However, as
there are no symmetry breaking terms in the quantum evolution,
the initial symmetry of the system being composed of a symmetric
photon and a symmetric mirror state
$\Omega_i~=~\Phi_i~\otimes~\Psi_i$ - must be conserved. Hence, one
may write the final wave function of the system as

\bea
\Omega_f~=~\sum{\Phi_s~\otimes~\Psi_s~+~\Phi_{as}~\otimes~\Psi_{as}}
\eea
where `s' and `as' stand for symmetric and anti-symmetric,
and the summation is over all possible foil states.

\begin{figure}
    \begin{center}\hspace{0mm}\mbox{\input epsf
\epsfxsize\columnwidth\epsfbox{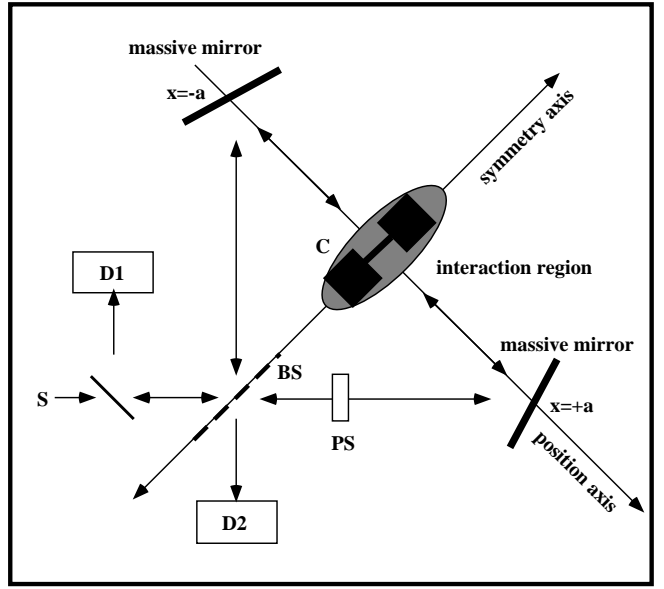}}\end{center}
\caption{The
open loop interferometer. Here, we position a two sided foil
mirror, held in an harmonic potential, into the interaction
region C, where the split incoming photon wave function "hits"
the foil from both sides. The beam splitter and the phase shifter
(with a phase equal to the phase difference between the reflected
and transmitted wave at the beam splitter), form a preparation system 
which ensures that the photon wave function is symmetric with respect
to the symmetry axis. The same system acts as a detection system with
parity eigen states.
Taking the mirror to be part of the set
up, one may say that we entangle the transversing photon to the
state of the set-up, which is also in a well defined quantum
state with a well defined spatial uncertainty. Using Parity
eigen-states to measure the state of the outgoing photon, we
learn about the decoherence of the set-up, without affecting it.
Thus, we are able to examine the quantum evolution of a
macroscopic system.\label{open}}
\end{figure}\noindent

Indeed as the parity measurement is in the above basis, it will
in fact collapse the mirror, which is entangled to the photon,
into a specific symmetry. In addition, measuring the energy of
the outgoing photon would collapse the mirror into {\em one}
specific energy eigen state of the harmonic oscillator. We post
select into our data set only photons for which the energy
difference $\Delta E$ between the incoming and outgoing photon
obeys $\Delta E=2m \hbar\omega$, where $m$ is a natural integer.
According to standard quantum evolution, there should be no
parity $-1$ (i.e. anti symmetric) detector clicks in our data set,
as the latter are, in the absence of localization, correlated to
a foil excitation of an odd number of levels.
Namely, within our post selected data set, detector D2 should remain dark.
Consequently, any such events in our data set 
must be a result of a non-standard
evolution of our foil, where through coupling to the environment
or through non quantum decoherence processes such as the
GRW mechanism \cite{Sq}. We end here our brief summary
of the proposed experiment, where a clear shortcoming of the
original discussion was that of being limited to initial pure
states, which are hard to experimentally achieve for macroscopic
objects.

Let us now start with an initial thermal state. Following the
object discussed in the original proposal, we begin by describing
the initial density matrix of a foil mirror oscillating in an
harmonic potential. Our basis is that of the energy eigen states,
where for convenience we limit the infinite basis to $N$ states
where $N\hbar\omega>>K_BT$, and where $T$ is the initial
temperature of the mirror.

\[%
\begin{array}
[c]{ccccccc}%
\rho(0)=\left(
\begin{array}
[c]{lllllll}%
p_{N-1} & 0 & . & . & . & 0 & 0\\
0 & p_{N-2} &   &   &   &   &  \\
. &   & . &   &   &   &  \\
. &   &   & . &   &   &  \\
. &   &   &   & . &   &  \\
0 &   &   &   &   & p_1 & 0\\
0 &   &   &   &   & 0 & p_0\\
\end{array}
\right)  ,
\end{array}
\]

where $p_n$ is simply the Boltzmann population
$e^{-(\frac{1}{2}+n)\hbar\omega/K_BT}$.

As $\rho(t)$ can always be looked upon as the sum of the density
matrices of each possible state i.e. $\rho(t)=\sum{l_k\rho_k(t)}$
where $l_k$ is determined by $\rho(0)=\sum{l_k\rho_k(0)}$ and is
simply the above Boltzmann factor - we examine in the following
the evolution of one specific pure state, and will afterwards sum.
Namely, we will proceed along the following path

\bea
<P_p>=\frac{Tr(P\rho_{red})}{Tr(\rho_{red})}=\frac{Tr(P(Tr_m\rho))}{Tr(Tr_m\rho)}=\nonumber\\
\frac{Tr(P(Tr_m\sum l_k\rho_k))}{Tr(Tr_m\sum
l_k\rho_k)}=\frac{\sum l_k(Tr(P(Tr_m\rho_k)))}{\sum l_k
Tr(Tr_m\rho_k)} \label{parity} \eea

where $P_p$ is the parity measurement on the outgoing photon,
$\rho_{red}$ is the reduced density matrix after a partial trace
$Tr_m$ on the mirror degrees of freedom, and $P$ in the energy
basis of the outgoing photon is (see Appendix A)

\[%
\begin{array}
[c]{ccccccc}%
P=\left(
\begin{array}
[c]{lllllll}%
1 & 0 & . & . & . & 0 & 0\\
0 & -1 &   &   &   &   &  \\
. &   & . &   &   &   &  \\
. &   &   & (-1)^n &   &   &  \\
. &   &   &   & . &   &  \\
0 &   &   &   &   & -1 & 0\\
0 &   &   &   &   & 0 & 1\\
\end{array}
\right)  ,
\end{array}
\]

and where that basis is $|j=-N>,|-N+1>,....,|N-1>,|N>$ where the
state $|j>$ means how many harmonic oscillator levels the mirror
has jumped due to the interaction with the photon. Note that
de-excitations of the mirror are in principle also valid.
Obviously for an even $j$, the above parity operator acting on
the $|j><j|$ matrix element, would return a positive number. In
the context of symmetric $|S>$ and anti-symmetric $|AS>$ outgoing
photons and taking $N$ to be even, the above basis can be written
as $|S_{-N}>,|AS_{-N+1}>,....,|AS_{N-1}>,|S_N>$.

It was shown in the original proposal that for any initial mirror
state $|n>$, the final state of the mirror-photon system is

\bea |\Omega^n_f\rangle=2\cos(k^{\Delta}x)|n\rangle_m
\frac{1}{\sqrt{2}}(|1\rangle|0\rangle+|0\rangle|1\rangle)_p+\nonumber\\
2i\sin(k^{\Delta}x)|n\rangle_m
\frac{1}{\sqrt{2}}(|1\rangle|0\rangle-|0\rangle|1\rangle)_p \eea

where $k^{\Delta}$ is simply the vectorial sum
$k^{\Delta}=k1+k2$, where $k1$ and $k2$ are the momenta of the
incoming and outgoing photon respectively.

As our energy post selection does not allow for odd excitations,
the remaining final state in our data set is

\bea |\Omega^n_f\rangle=2\cos(k^{\Delta}x)|n\rangle_m
\frac{1}{\sqrt{2}}(|1\rangle|0\rangle+|0\rangle|1\rangle)_p=\nonumber\\
\alpha_0|0>_m|S_{-n}>_p+\alpha_2|2>_m|S_{-n+2}>_p+.... \eea

Indeed, it is already clear at this stage that for any given
initial pure state our data set will only include parity $+1$
photons, which means our apparatus will be sensitive to the
symmetry breaking expected to appear in the form of anti
symmetric photons. A summation over many such initial states that
form the initial thermal state would not change this situation.
However, let us arrive at this conclusion in a formal way by
tracing over the degrees of freedom of the mirror to arrive at
the result of a parity measurement on the outgoing photon.

Taking into account all the above, the density matrix $\rho_k$
after the mirror-photon interaction for any initial pure state
$|n>$ has the form

\bea \rho_n=|\Omega^n_f><\Omega^n_f|=
|\alpha_0|^2|S_{-n}>|0><0|<S_{-n}|+\nonumber\\
|\alpha_2|^2|S_{-n+2}>|2><2|<S_{-n+2}|+....\eea

where

\bea
|\alpha_m|^2=P_{n\rightarrow m(even)}=|\langle
n|\cos(k^{\Delta}x)|m\rangle|^2
\eea

The partial trace now simply gives

\bea G=Tr_m(\rho_n)=\nonumber\\
<0|\rho_n|0>+<1|\rho_n|1>+<2|\rho_n|2>+...=\nonumber\\
|\alpha_0|^2|S_{-n}><S_{-n}|+|\alpha_2|^2|S_{-n+2}><S_{-n+2}|+....
\eea

As only the even positions on the diagonal of the photon density
matrix have non zero element, it is clear that $PG=G$. Hence from
eq. \ref{parity} we find

\bea 
<P_p>=\frac{\sum l_n(Tr(P(G)))}{\sum l_n Tr(G)}=+1 
\eea

\begin{figure}
    \begin{center}\hspace{0mm}\mbox{\input epsf
\epsfxsize\columnwidth\epsfbox{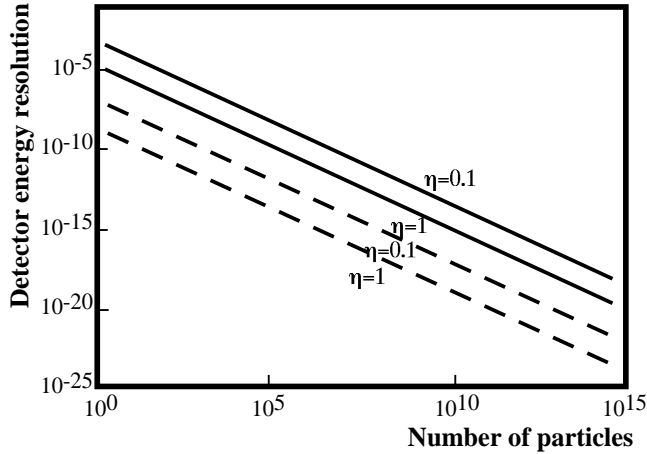}}\end{center}
\caption{Needed energy resolution $\frac{\Delta E}{E_p}$ as a
function of the number of foil nucleons, for two extremes of the
feasibly used light spectrum: x-ray and red (in dash). $E_p$ is the
photon energy while $\Delta E=\hbar \omega$.
As an example, two $\eta$ values are given, where $\eta$ is the
Lamb-Dicke parameter equal to the ground state size divided by the 
light wave length $\lambda$ i.e. 
$\frac{4\pi}{\lambda}\sqrt{\hbar/2m\omega}$
\label{pres}}
\end{figure}\noindent

Let us finalize this short letter by looking at the needed energy 
detection resolution. In figure \ref{pres} we plot the needed resolution
as a function of the mass of the foil. Obviously, if one wants to maintain
a reasonable ground state size (see implications in ref. \cite{folman}),
as the oscillating mass is increased the frequency has to be decreased and
hence the energy level specing becomes smaller. Consequently, the needed
energy detection resolution becomes more stringent.  

We see that for currently available resolutions of $10^{-13}$, one
can already perform the experiment with a $10^9$ particle mirror.

To conclude, we have shown that as long as the energy resolution
of our detectors allows us to ignore all odd excitations or
relaxations of the mirror, all out going photons should be
symmetric and consequently click only in detector D1. This allows
us to be sensitive to symmetry breaking events originating from
decoherence processes, even if initially the mirror was in a
thermal state.
\\

Acknowledgement\\

This work was supported by Marie Curie fellowship number HPMF CT
1999-00235.

\appendix

\section{Energy basis parity operator}

The parity operator in the descrete position space (e.g. the 2
dimensional double slit experiment) is well known and is presented in
the introduction. In infinite space, the parity operator may be written as

\bea
P_x=\int |-x'><x'|dx'
\eea

with $|x>$ being the position eigen state at point x.

As the energy eigen state $|n>$ may be written as $\int \psi_n(x) |x> dx$,
where $\psi_n(x)=<n|x>$ is the wave function of $|n>$, we have

\bea
P_x|n>=\int |-x'><x'| |n>dx'=\int |-x'> \psi_n(x')dx'=\nonumber\\
\int |x'> \psi_n(-x')dx'=\int (-1)^n \psi_n(x')|x'>dx'=\nonumber\\
(-1)^n |n> 
\eea

Namely, the eigen values of the parity operator applied on the energy
eigen modes $|n>$ are $\lambda_n=(-1)^n$, which means that in this basis, 
the operator takes a diagonal form with its diagonal matrix elements equal 
to $\lambda_n$, which is what we wanted to prove.


\vspace{-5mm}
\begin{references}
\vspace{-15mm}

\bibitem{dec}
    D. Guilini et al., {\em Decoherence and the appearance of a classical
world in quantum theory}, Springer (1996), and references therein.

\bibitem{greenberger}
    D.M. Greenberger \& A. Yasin, Phys. Lett. A {\em 128}, 391 (1988);
W.K. Wooters \& W.H. Zurek, Phys. Rev. D {\em 19}, 473 (1979).

\bibitem{englert}
    B.G. Englert, Phys. Rev. Lett. {\em 77}, 2154 (1996); 
P.D.D. Schwindt et al., Phys. Rev. A {\em 60}, 4285 (1999).

\bibitem{folman}
    R. Folman et al., Eur. Phys. J. D {\em 13}, 93 (2001).

\bibitem{Sq}
    P. Pearle \& E. Squires, Phys. Rev. Lett. {\em 73}, 1 (1994) 
and references therein.

\end{references}
\end{document}